\documentclass[
aps,
pra,
reprint,
floatfix,
amsmath,
twocolumn,
nofootinbib,
superscriptaddress
]{revtex4-1}

\usepackage{dcolumn}   
\usepackage{bm}        
\usepackage{amssymb}   
\usepackage{amsfonts}
\usepackage{color}
\usepackage{lineno}
\usepackage{verbatim}
\usepackage{enumitem}
\usepackage{mathrsfs}
\usepackage{graphicx}
\usepackage{amsmath}
\usepackage{amsfonts}
\usepackage{ulem}
\usepackage[]{hyperref}
\usepackage{xcolor}

\hypersetup{
	colorlinks,
	linkcolor={blue!40!black},
	urlcolor={blue!40!black},
	citecolor={blue!40!black}
}
\usepackage{etoolbox}
\makeatletter
\pretocmd{\NAT@open}{\begingroup\color{\@citecolor}}{}{}
\apptocmd{\NAT@close}{\endgroup}{}{}
\makeatother

\usepackage{makecell}
\newcommand*\chem[1]{\ensuremath{\mathrm{#1}}}

\def\lsim{\mathrel{\rlap{\lower4pt\hbox{\hskip1pt$\sim$}}
    \raise1pt\hbox{$<$}}}                
\def\gsim{\mathrel{\rlap{\lower4pt\hbox{\hskip1pt$\sim$}}
    \raise1pt\hbox{$>$}}}                

\usepackage{titlesec}
\titlespacing\section{0pt}{22pt plus 4pt minus 2pt}{11pt plus 2pt minus 2pt}
\titlespacing\subsection{0pt}{22pt plus 4pt minus 2pt}{10pt plus 2pt minus 2pt}

\begin{document}
\normalem

\title{Digital coherent control of a superconducting qubit}

\author{E. Leonard Jr.}
\thanks{These authors contributed equally to this work.}
\author{M. A. Beck}
\thanks{These authors contributed equally to this work.}
\affiliation{Department of Physics, University of Wisconsin-Madison, Madison, Wisconsin 53706, USA}
\author{J. Nelson}
\affiliation{Department of Physics, Syracuse University, Syracuse, NY 13244, USA}

\author{B. G. Christensen}
\author{T. Thorbeck}
\thanks{Present address: IBM T. J. Watson Research Center, Yorktown Heights, New York 10598, USA}
\affiliation{Department of Physics, University of Wisconsin-Madison, Madison, Wisconsin 53706, USA}
\author{C. Howington}
\affiliation{Department of Physics, Syracuse University, Syracuse, NY 13244, USA}
\author{A. Opremcak}
\author{I. V. Pechenezhskiy}
\thanks{Present address: University of Maryland, College Park, Maryland 20742, USA}
\affiliation{Department of Physics, University of Wisconsin-Madison, Madison, Wisconsin 53706, USA}
\author{K. Dodge}
\affiliation{Department of Physics, Syracuse University, Syracuse, NY 13244, USA}
\author{N. P. Dupuis}
\affiliation{Department of Physics, University of Wisconsin-Madison, Madison, Wisconsin 53706, USA}
\author{J. Ku}
\affiliation{Department of Physics, Syracuse University, Syracuse, NY 13244, USA}
\author{F. Schlenker}
\author{J. Suttle}
\author{C. Wilen}
\author{S. Zhu}
\affiliation{Department of Physics, University of Wisconsin-Madison, Madison, Wisconsin 53706, USA}

\author{M. G. Vavilov}
\affiliation{Department of Physics, University of Wisconsin-Madison, Madison, Wisconsin 53706, USA}
\author{B. L. T. Plourde}
\affiliation{Department of Physics, Syracuse University, Syracuse, NY 13244, USA}

\author{R. McDermott}
\thanks{electronic address: \url{rfmcdermott@wisc.edu}}
\affiliation{Department of Physics, University of Wisconsin-Madison, Madison, Wisconsin 53706, USA}

\date{\today}


\begin{abstract}

High-fidelity gate operations are essential to the realization of a fault-tolerant quantum computer. In addition, the physical resources required to implement gates must scale efficiently with system size. A longstanding goal of the superconducting qubit community is the tight integration of a superconducting quantum circuit with a proximal classical cryogenic control system. Here we implement coherent control of a superconducting transmon qubit using a Single Flux Quantum (SFQ) pulse driver cofabricated on the qubit chip. The pulse driver delivers trains of quantized flux pulses to the qubit through a weak capacitive coupling; coherent rotations of the qubit state are realized when the pulse-to-pulse timing is matched to a multiple of the qubit oscillation period. We measure the fidelity of SFQ-based gates to be $\sim$95\% using interleaved randomized benchmarking. Gate fidelities are limited by quasiparticle generation in the dissipative SFQ driver. We characterize the dissipative and dispersive contributions of the quasiparticle admittance and discuss mitigation strategies to suppress quasiparticle poisoning. These results open the door to integration of large-scale superconducting qubit arrays with SFQ control elements for low-latency feedback and stabilization.

\end{abstract}

\maketitle 


\section{INTRODUCTION}\label{sec:int}

Josephson qubits are a promising candidate for the construction of a large-scale quantum processor~\cite{Clarke:2008aa,Devoret:2013aa,Gambetta:2017aa}. Gate and measurement fidelities have surpassed the threshold for fault-tolerant operations in the two-dimensional surface code~\cite{Fowler:2012aa,Barends:2014aa}, and the successful demonstration of small repetition codes~\cite{Riste:2015aa,Kelly:2015aa} provides a direct validation of the fidelity of the quantum hardware and control and measurement elements that will be needed for a first demonstration of error-protected logical qubits. Current approaches to the control of superconducting qubits rely on the application of pulsed microwave signals. The generation and routing of these control pulses involve substantial experimental overhead in the form of room-temperature and cryogenic electronics hardware. This includes, but is not limited to, coherent microwave sources, arbitrary waveform generators, quadrature mixers, and amplifiers, as well as coaxial lines and signal conditioning elements required to transmit these signals into the low-temperature experimental environment.

While brute-force scaling of current technology may work for moderate-sized superconducting qubit arrays comprising hundreds of devices, the control of large-scale systems of thousands or more qubits will require fundamentally new approaches. An attractive candidate for the control of large-scale qubit arrays is the Single Flux Quantum (SFQ) digital logic family~\cite{Likharev:1991aa}. In SFQ digital logic, classical bits of information are encoded in fluxons, propagating voltage pulses whose time integral is precisely quantized to $h/2e\equiv\Phi_0$, the superconducting flux quantum. Here, the presence (absence) of a quantized flux pulse across a Josephson junction (JJ) in the SFQ circuit during a given clock cycle constitutes the classical bit ``1" (``0"). There have been prior attempts to integrate SFQ digital logic with superconducting qubits~\cite{Zhou:2001aa,Crankshaw:2003aa,Semenov:2003aa}, with some notable recent successes in the area of qubit measurement~\cite{Fedorov:2014aa}. The work presented here is motivated by a previous study which showed that resonant trains of SFQ pulses can be used to induce high-fidelity qubit rotations, with gate fidelity limited by leakage errors~\cite{McDermott:2014aa}. Subsequent work based on optimal control theory provided a proof-of-principle demonstration that leakage errors can be suppressed significantly by clocking control bits at a higher frequency and allowing variation in the pulse-to-pulse timing intervals~\cite{Liebermann:2016aa}.
\begin{figure}[t]
\includegraphics{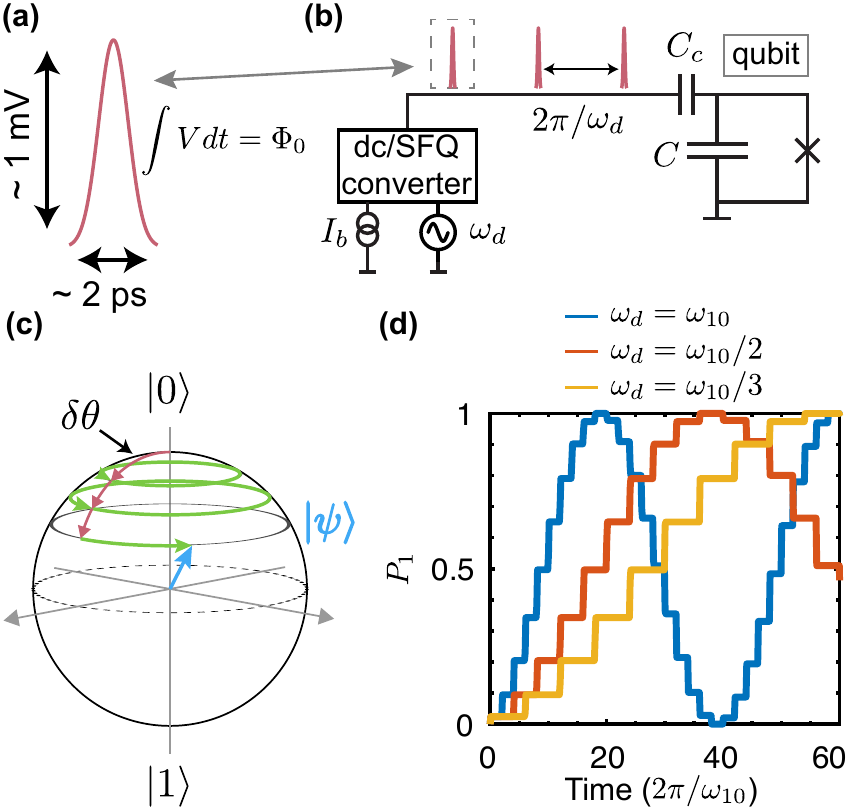}
\caption{\label{fig:cartoon} (Color) Coherent control of a qubit using SFQ pulses. \textbf{(a)} A phase slip across a JJ induces a voltage pulse with time integral precisely quantized to $h/2e\equiv\Phi_0\approx2.07$~mV$\times$ps, the superconducting flux quantum. For typical parameters, pulse amplitudes are of order 1~mV and pulse durations are of order 2~ps. \textbf{(b)} Block diagram of the experiment. An external microwave tone with frequency $\omega_d$ is used to trigger a dc/SFQ converter, which generates a train of pulses with interpulse timing of $2\pi/\omega_d$; the resulting pulse train is coupled capacitively to a transmon qubit. \textbf{(c)} Trajectory of the qubit state vector on the Bloch sphere due to a resonant train of SFQ pulses. Each SFQ pulse induces an incremental rotation; the green parallels represent qubit free precession in between SFQ pulses. \textbf{(d)} Simulated qubit Rabi oscillations for SFQ drive at various subharmonics of the qubit transition frequency $\omega_{10}$. Discrete steps in excited state population $P_1(t)$ are induced by the arrival of single SFQ pulses; in between pulses, the qubit undergoes free precession with no change in $P_1(t)$. In \textbf{(c)} and \textbf{(d)}, a large, non-optimal value of $\delta\theta=\pi/10$ is chosen for the purpose of display.}
\end{figure}

In this manuscript, we describe the arbitrary coherent control of a superconducting transmon qubit driven by SFQ pulses. The SFQ pulses are generated by a dc/SFQ converter~\cite{Likharev:1991aa} cofabricated on the same chip as the qubit. The basic scheme of the experiment is shown in Fig.~\ref{fig:cartoon}\hyperref[fig:cartoon]{(a,~b)}. The SFQ driver circuit is biased with a dc current and an external microwave tone is applied to the trigger port of the driver. When the trigger tone exceeds a certain threshold, SFQ pulses are generated and coupled capacitively to the qubit. Each pulse induces an incremental rotation on the Bloch sphere~\cite{McDermott:2014aa}
\begin{align}
\delta\theta&=C_c \Phi_0 \sqrt{\frac{2\omega_{10}}{\hbar C}},
\label{eqn:dtheta}
\end{align}
where $C_c$ is the coupling capacitance between the SFQ driver and qubit, $C$ is the qubit self-capacitance, and $\omega_{10}$ is the qubit fundamental transition frequency [see Fig.~\ref{fig:cartoon}\hyperref[fig:cartoon]{(c)}]. When the pulse-to-pulse spacing is matched to an integer multiple of the qubit precession period, the state vector undergoes a coherent rotation on the Bloch sphere about a control vector whose direction is determined by the relative timing of the pulse sequence. In Fig.~\ref{fig:cartoon}\hyperref[fig:cartoon]{(d)} we show simulated SFQ-induced Rabi oscillations for SFQ pulse trains resonant with $\omega_{10}$ as well as subharmonics $\omega_{10}/2$ and $\omega_{10}/3$; here, for the purposes of display we assume a large, non-optimal, per-pulse rotation $\delta \theta = \pi/10$. A related idea for selective excitation was pursued by some early practitioners of nuclear magnetic resonance, who termed the pulse sequence DANTE (Delays Alternating with Nutations for Tailored Excitation), as the terraced trajectory of the state vector on the Bloch sphere recalls the multilevel structure of the \emph{Inferno}~\cite{Bodenhausen:1976aa,Morris:1978aa}.

This manuscript is organized as follows. In Section~\ref{sec:ic}, we describe the cofabrication of the SFQ pulse driver and transmon qubit on a single chip and discuss separate characterization of the driver and qubit circuits. In Section~\ref{sec:gates}, we discuss implementation of an orthogonal gate set using SFQ pulses and present the results of gate characterization using interleaved randomized benchmarking (RB)~\cite{Magesan:2012aa}; the achieved gate fidelities $\mathcal{F}\sim 95\%$ are limited by quasiparticle (QP) poisoning of the qubit induced by operation of the dissipative SFQ pulse driver. In Section~\ref{sec:qp}, we discuss experiments to characterize the real and imaginary parts of the QP admittance seen by the qubit and explore the dynamics of QP generation and relaxation in the circuit. Finally, in Section~\ref{sec:conc} we discuss straightforward improvements that should provide a significant suppression of QP poisoning in next-generation systems, and we comment on the prospects for high-fidelity control of a large-scale multiqubit circuit using a proximal SFQ-based pulse programmer.


\section{THE QUANTUM-CLASSICAL HYBRID~CIRCUIT\label{sec:ic}}

\begin{figure*}[t]
\includegraphics{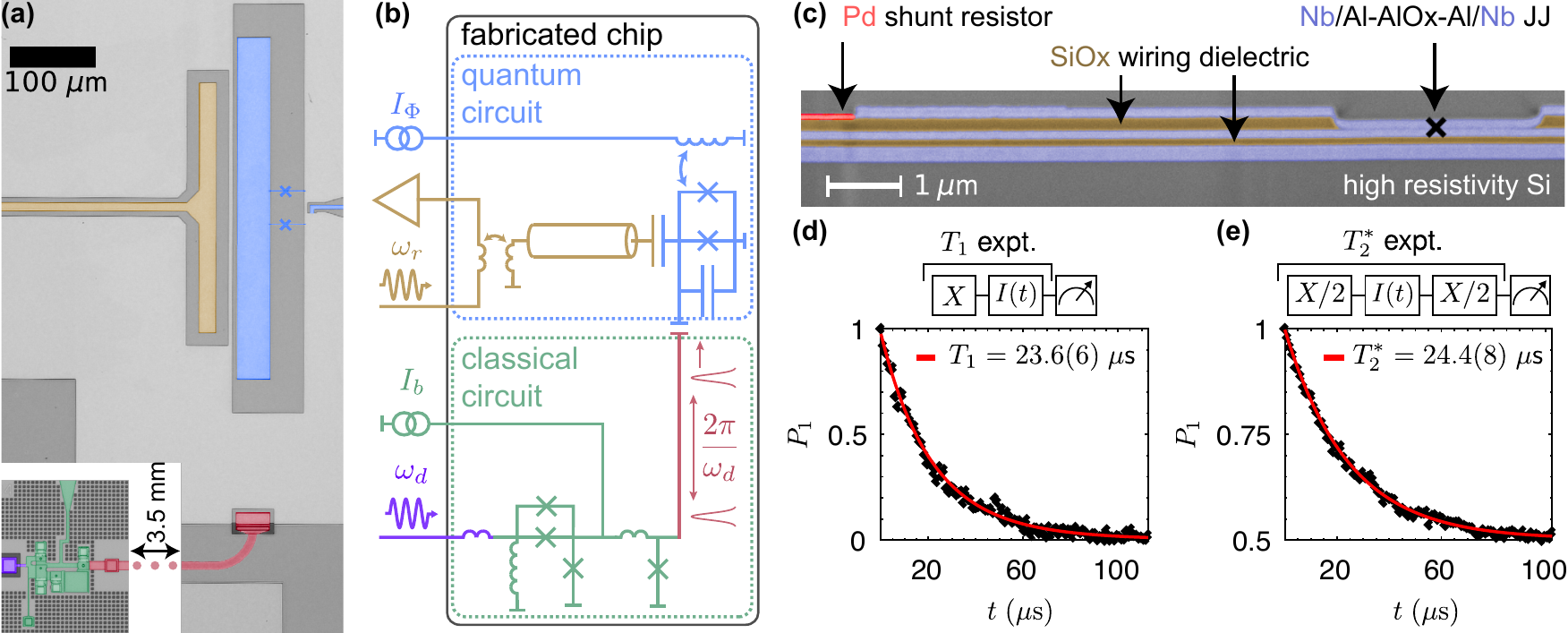}
\caption{\label{fig:overview} (Color) Overview of the quantum-classical hybrid circuit. \textbf{(a)} False-color optical micrograph of the SFQ driver circuit (green) with capacitive coupling (red) to the superconducting transmon qubit (blue). The qubit is coupled to the voltage antinode of a $\lambda/4$ coplanar waveguide resonator (yellow) and is biased by a dc flux line (blue). The SFQ driver is triggered by a microwave tone delivered via a $Z=50\;\Omega$ CPW transmission line (purple).  The distance between the SFQ driver and qubit is approximately 3.5~mm. \textbf{(b)} Circuit diagram for the device of \textbf{(a)}. Each JJ in the classical circuit is shunted by a thin-film Pd resistor (not shown). \textbf{(c)} False-color cross-sectional SEM micrograph showing layer stack in the vicinity of a Nb/Al-\chem{AlO_x}-Al/Nb JJ of the SFQ driver. The Nb wiring layers (blue) are separated by \chem{SiO_x} dielectric layers (brown); the driver junctions are formed in vias in the dielectric layer separating the second and third Nb metal layers. Pd resistors (red) shunt the junctions of the driver circuit. \textbf{(d-e)} Conventional microwave-based characterization of qubit energy relaxation~\textbf{(d)} and dephasing~\textbf{(e)} with extracted characteristic times $T_1=23.6(6)\;\mu$s and $T_2^*=24.4(8)\;\mu$s, respectively.}
\end{figure*}

The experiment involves a transmon qubit cofabricated with an SFQ driver circuit on a high-resistivity Si(100) substrate. In Fig.~\ref{fig:overview}\hyperref[fig:overview]{(a)} we show a micrograph of the completed circuit, and in Fig.~\ref{fig:overview}\hyperref[fig:overview]{(b)} we present the circuit diagram. The transmon qubit [blue in Fig.~\ref{fig:overview}\hyperref[fig:overview]{(a,~b)}] includes a flux-tunable compound JJ with asymmetry of approximately 2:1~\cite{Hutchings:2017aa}. For the experiments described here, the qubit is tuned to its upper flux-insensitive sweet spot with fundamental transition frequency of $\omega_{10}/2\pi=4.958$~GHz; the qubit anharmonicity is $\alpha/2\pi\approx-220$~MHz. The qubit is independently addressable with either shaped microwave tones or SFQ pulses and is measured via standard heterodyne detection through a $\lambda/4$ coplanar waveguide (CPW) resonator (yellow) with a resonance frequency of $6.15$~GHz, decay rate $\kappa=(500\:\textrm{ns})^{-1}$, and resonator-qubit coupling of $g/2\pi\sim100$~MHz. The SFQ driver circuit [green in Fig.~\ref{fig:overview}\hyperref[fig:overview]{(a,~b)}] is based on a standard dc/SFQ converter~\cite{Likharev:1991aa}. The output of the driver is coupled via a single-cell Josephson transmission line to a short superconducting microstrip with characteristic impedance $\sim 1\;\Omega$; the microstrip is terminated in a coupling capacitance $C_c$ [red in Fig.~\ref{fig:overview}\hyperref[fig:overview]{(a,~b)}], which delivers SFQ pulses to the qubit. The SFQ driver circuit is biased with a dc current $I_b$ (green) and an external microwave trigger tone at frequency $\omega_d$ (purple); a properly biased circuit will output a single SFQ pulse for each cycle of the trigger waveform. We summarize the key fabrication steps used to realize the hybrid circuit below, with further detail included in Appendix~\ref{app:fab}.

\subsection{Device Fabrication\label{sec:fab}}

The qubit capacitor, CPW readout resonator, and surrounding microwave ground plane are realized in a 180~nm-thick Nb film grown by dc magnetron sputtering. The dielectric materials required for the multilayer stack are deposited using plasma enhanced chemical vapor deposition (PECVD). The SFQ driver circuit incorporates three superconducting Nb layers, two insulating \chem{SiO_x} layers, one normal metal Pd layer, and four Nb/Al-\chem{AlO_x}-Al/Nb JJs with critical current density $J_c\sim1$~kA/cm$^2$ and areas $\{3,4,4,5.5\}\mbox{ }\mu\mbox{m}^2$. The SFQ driver JJs are shunted by thin-film Pd resistors formed by electron-beam evaporation and liftoff. In areas where good metallic contact is required between wiring layers, an \textit{in situ} Ar ion mill cleaning step is used to remove native \chem{NbO_x} prior to metal deposition. Figure~\ref{fig:overview}\hyperref[fig:overview]{(c)} shows a cross-sectional scanning electron microscope (SEM) image of the layer stack in the vicinity of the SFQ driver obtained following a focused ion beam mill through the circuit. Here, the Nb layers are false-colored in blue, the \chem{SiO_x} dielectric layers are colored brown, and the Pd shunt resistor is shown in red. A black $\mathbb{\times}$ marks the location of a Nb/Al-\chem{AlO_x}-Al/Nb junction interface.

Throughout fabrication of the multilayer SFQ driver, a blanket layer of \chem{SiO_x} is used to protect the areas where the readout resonator and qubit capacitor and junctions will be formed. Following completion of the SFQ driver circuit, this dielectric protection layer is removed and the quantum circuit is patterned and etched. Formation of the qubit junctions is the final step of the process. The transmon junction electrodes are defined using electron-beam lithography with a standard MMA/PMMA resist bilayer. The qubit junctions are grown via double-angle Al evaporation with a controlled \textit{in situ} oxidation between the two deposition steps.

\subsection{Experimental Setup\label{sec:expt}}

\begin{figure*}[t]
\includegraphics{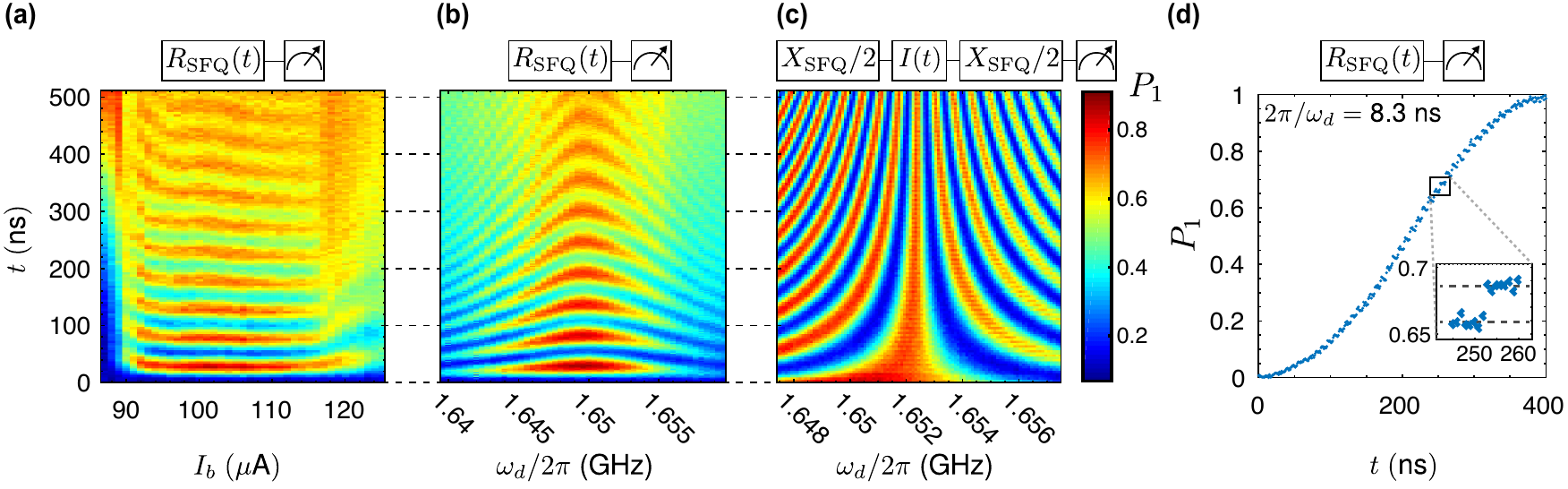}
\caption{\label{fig:rabi} (Color) Basic qubit operations driven by SFQ pulses.  \textbf{(a)} SFQ-based Rabi oscillations as a function of bias current $I_b$ to the SFQ driver circuit. Here, $\omega_d=\omega_{10}/3\approx1.65$~GHz. \textbf{(b)} Rabi chevron experiment and \textbf{(c)} Ramsey fringe experiment where the SFQ pulse frequency $\omega_d$ is varied slightly in the vicinity of $\omega_{10}/3$.  \textbf{(d)} (main) Time trace of a single SFQ-based Rabi flop obtained with a pulse rate $\omega_{10}/41=120.90$~MHz. (inset) A zoom-in on discrete steps in $P_1(t)$ occurring every $8.3$~ns, the SFQ pulse-to-pulse timing interval.}
\end{figure*}

The device is wirebonded in an SMA-connectorized Al package and loaded on the experimental stage of a dilution refrigerator. Low-bandwidth coaxial cabling is used for the current bias $I_b$ of the SFQ driver and the flux bias $I_{\Phi}$ of the transmon qubit. These signals are generated by isolated low-noise sources conditioned by several layers of low-pass filtering at the 4~K and millikelvin stages. Microwave pulses for control and readout are derived by single-sideband mixing of local oscillator (LO) tones with intermediate frequency (IF) signals generated by dual-channel 1 GS/s arbitrary waveform generators. These high-frequency signals are attenuated and conditioned at various stages of the measurement cryostat. A full wiring diagram of the experimental setup can be found in Appendix~\ref{app:meas}.

Prior to investigation of SFQ-based gates, the transmon qubit and SFQ pulse driver are characterized separately. The qubit is flux tuned to its upper sweet spot to reduce susceptibility to flux noise~\cite{Hutchings:2017aa}. Microwave excitation pulses are coupled to the qubit via off-resonant drive through the readout resonator. Standard inversion recovery and Ramsey sequences are used to determine the qubit energy relaxation and dephasing times $T_1$ and $T_2^*$, respectively; data from these experiments are shown in Fig.~\ref{fig:overview}\hyperref[fig:overview]{(d,~e)}. We find relaxation times in excess of 20~$\mu$s. These times are compatible with relaxation times that are achieved in planar transmon qubits fabricated using a single double-angle evaporation step that avoids lossy dielectrics~\cite{Devoret:2013aa}.

Bringup of the SFQ driver circuit involves measurement of the device $IV$-characteristic curves. In the absence of applied microwave drive at the trigger input, the critical current $I_c$ of the device is around 130~$\mu$A. Proper operation of the SFQ driver is determined by biasing $I_b$ near the critical current and applying microwave drive to the trigger input. For appropriate drive amplitude, the application of an oscillatory tone at frequency $\omega_d$ induces a Shapiro step in the driver $IV$ curve at voltage $\Phi_0 \omega_d/2\pi$~\cite{Shapiro:1963aa}.


\section{SFQ PULSES FOR QUBIT~CONTROL\label{sec:gates}}

Following establishment of coarse operating points for the qubit and SFQ driver, the qubit itself is used to fine-tune SFQ-based gate operations. We apply a microwave tone at a subharmonic of the qubit frequency $\omega_d=\omega_{10}/n$, where $n\in\{2,3,4,\dots\}$, to the driver trigger input for varying time and sweep the converter bias $I_b\lesssim I_c$ while monitoring the state of the qubit [Fig.~\ref{fig:rabi}\hyperref[fig:rabi]{(a)}]. In this device, subharmonic drive is essential in order to circumvent direct microwave crosstalk from the trigger line to the qubit. For a broad range of driver bias current, the qubit Rabi frequency is approximately independent of driver bias. Similarly, the Rabi frequency is independent of microwave trigger drive power over a broad range (not shown); this is expected, as the SFQ driver acts as a threshold comparator generating a single SFQ pulse per period of the trigger waveform. From this measurement, we select the optimal current bias and microwave trigger power to produce clear qubit Rabi oscillations.

At this point, we use SFQ pulse trains to perform conventional qubit characterization. In Fig.~\ref{fig:rabi}\hyperref[fig:rabi]{(b)} we show the results of an SFQ-based Rabi experiment where the SFQ pulse frequency is swept over a narrow interval in the vicinity of $\omega_{10}/3$ to produce the familiar ``chevron'' interference pattern. From the data we determine the length of SFQ-based $\pi/2$ and $\pi$ gates, which we denote as $X_{\textrm{SFQ}}/2$ and $X_{\textrm {SFQ}}$, respectively. For $n=3$, the duration of the $X_{\textrm{SFQ}}/2$ ($X_{\textrm{SFQ}}$) gate is 14 (28)~ns, corresponding to 23 (46) discrete SFQ pulses. From the measured Rabi frequency and Eq.~\hyperref[eqn:dtheta]{(}\ref{eqn:dtheta}\hyperref[eqn:dtheta]{)} we determine the coupling capacitance $C_c=400$~aF between the driver and qubit circuits. Once the $X_{\textrm{SFQ}}/2$ gate is defined, we perform a Ramsey interferometry experiment [Fig.~\ref{fig:rabi}\hyperref[fig:rabi]{(c)}]; we obtain familiar interference fringes oscillating with a frequency corresponding to three times the detuning of the SFQ pulse train from $\omega_{10}/3$.

By triggering the SFQ driver at a much lower frequency and delivering a dilute SFQ pulse train to the qubit, we can map out the terraced trajectory of the qubit state vector on the Bloch sphere evoked by the name of the DANTE pulse sequence. In Fig.~\ref{fig:rabi}\hyperref[fig:rabi]{(d)} we show data from a qubit Rabi flop obtained with an SFQ pulse train triggered at a deep subharmonic $\omega_d=\omega_{10}/41$. The discrete steps induced by the individual SFQ pulses are evident and emphasized in the figure inset.

\begin{figure}[t]
\includegraphics{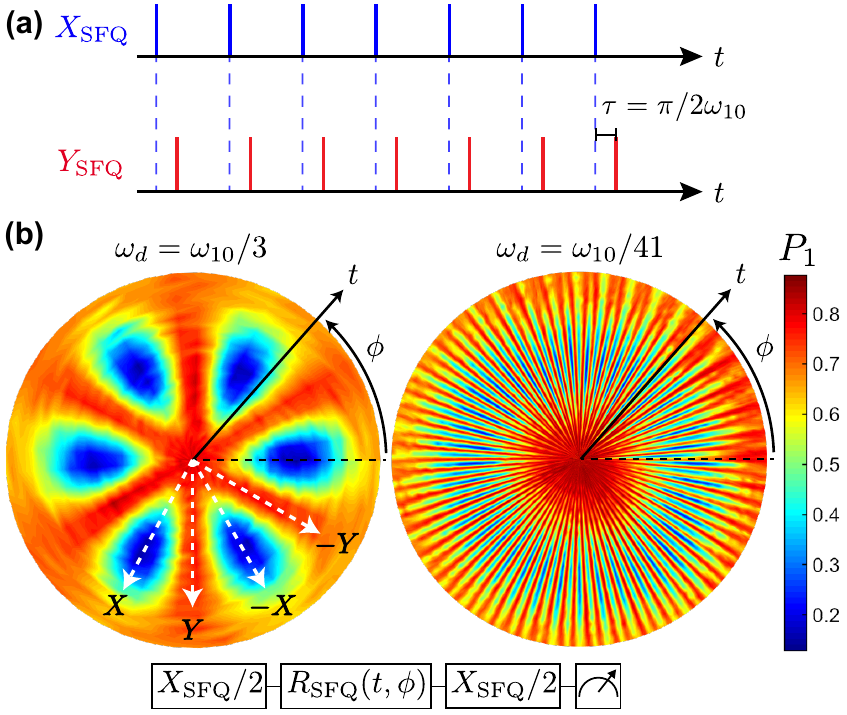}
\caption{\label{fig:ortho} (Color) \textbf{(a)} Illustration depicting SFQ pulse sequences for orthogonal qubit control. The resonant $X$-sequence establishes a timing reference; a change in the relative timing of a second resonant sequence can be used to access different orientations of the qubit control vector in the equatorial plane of the Bloch sphere. A $Y_{\textrm{SFQ}}$-sequence is shifted in time by one quarter of the qubit oscillation period with respect to the $X_{\textrm{SFQ}}$-sequence. \textbf{(b)} Generalized Rabi scans for resonant SFQ pulse trains triggered at subharmonic frequency $\omega_{10}/3$ (left) and $\omega_{10}/41$ (right), respectively. The pulse sequence is shown below the figure. The radial coordinate is swept by incrementing the duration of $R_{\textrm{SFQ}}(t,\phi)$ from 0~ns to 60~ns (left) or 660~ns (right), while the polar angle corresponds to the phase $\phi$ of the trigger waveform used to generate the pulse train. The length and phase of each $X_{\textrm{SFQ}}/2$ pulse are fixed. Adjustment of the phase $\phi$ of the trigger waveform by $\pi/2n$ provides access to an orthogonal control axis; here, $n$ is the subharmonic drive factor. The $2n$-fold symmetry of the plots is explained by the fact that subharmonic drive automatically yields an $n$-fold increase in the number of (time-shifted) resonant sequences that access the same control vector on the Bloch sphere, and by the fact that our sequence doesn't distinguish between positive and negative rotations (yielding an additional doubling of the symmetry of the plot). On the left panel, we label rays corresponding to trigger waveform phases that yield $\pm X_{\textrm{SFQ}}, \pm Y_{\textrm{SFQ}}$ rotations.}
\end{figure}

Any complete qubit control scheme requires rotations about orthogonal axes. In the case of conventional microwave-based qubit control, the direction of the rotation is set by the phase of the pulse. For SFQ-based control, the relative timing of two resonant SFQ pulse trains determines the directions of the corresponding control vectors in the equatorial plane of the Bloch sphere. We first consider the case of resonant drive at the qubit fundamental frequency. We arbitrarily choose the relative timing of one sequence to correspond to the $X_{\textrm{SFQ}}$-rotation. For a second resonant pulse sequence that is shifted in time by $\tau$ with respect to the first, the qubit sees a control vector that is rotated by an angle $\phi = \omega_{10}\tau$ in the equatorial plane. For example, the pulse train for the $Y_{\textrm{SFQ}}$-gate is shifted by an amount $\tau = \pi/2 \omega_{10}$ (one quarter of an oscillation period) with respect to that of the $X_{\textrm{SFQ}}$-gate; see Fig.~\ref{fig:ortho}\hyperref[fig:ortho]{(a)}. In practice, the relative timing between pulse trains $\tau$ is controlled by adjusting the phase of the SFQ trigger waveform using standard heterodyne techniques.

For SFQ drive at a subharmonic of the qubit frequency, timing shifts between sequences induce a faster rotation of the angle of the control vector by $\phi_n = n \omega_{10}\tau$, where $n$ is the subharmonic drive factor. Figures~\ref{fig:ortho}\hyperref[fig:ortho]{(b,~c)} display generalized 2D Rabi data for qubits driven with SFQ pulse trains at $\omega_{10}/3$ and $\omega_{10}/41$, respectively. A Rabi pulse with varying duration and phase is inserted between two SFQ-based $\pi/2$ gates whose timing defines the $X$-direction. In these plots, the duration $t$ of the control sequence is encoded in the radial direction, the polar coordinate is given by the phase of the SFQ trigger tone at frequency $\omega_d=\omega_{10}/n$, and the measured qubit population is plotted in false color. We see clear $2n$-fold symmetry in the scans: pulse trains shifted by $2\pi/n \omega_{10}$ are equivalent, while the additional factor of 2 comes from the fact that the sequence does not distinguish between positive and negative rotations. In the plots, polar angles that yield high-contrast Rabi oscillations correspond to control in the $\pm X$ directions, while polar angles that yield no oscillation correspond to orthogonal control in the $\pm Y_{\textrm{SFQ}}$ directions. 

\begin{figure}[b]
\includegraphics{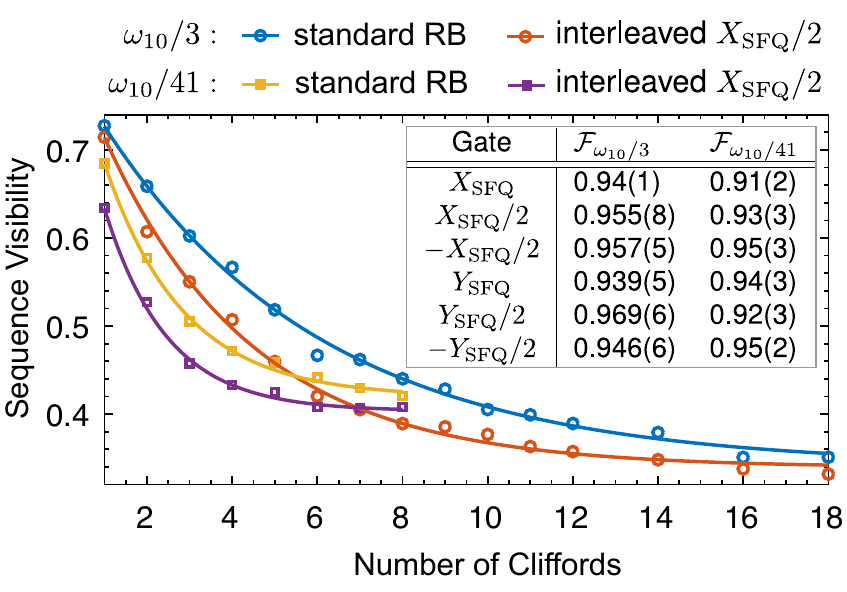}
\caption{\label{fig:rb} (Color) Randomized benchmarking (RB) of SFQ-based gates. (main) Depolarizing plot for all Cliffords and for the interleaved gate sequence designed to probe the $X_{\textrm{SFQ}}/2$ gate as a function of the number of Clifford operations. The (blue, orange) traces correspond to SFQ sequences with pulse rate $\omega_{10}/3$, while the (yellow, purple) traces correspond to pulse rate $\omega_{10}/41$. (inset) Table summarizing RB data for SFQ-based gates with $\omega_d=\omega_{10}/\{3,41\}$.}
\end{figure}

\begin{figure*}[t]
\includegraphics{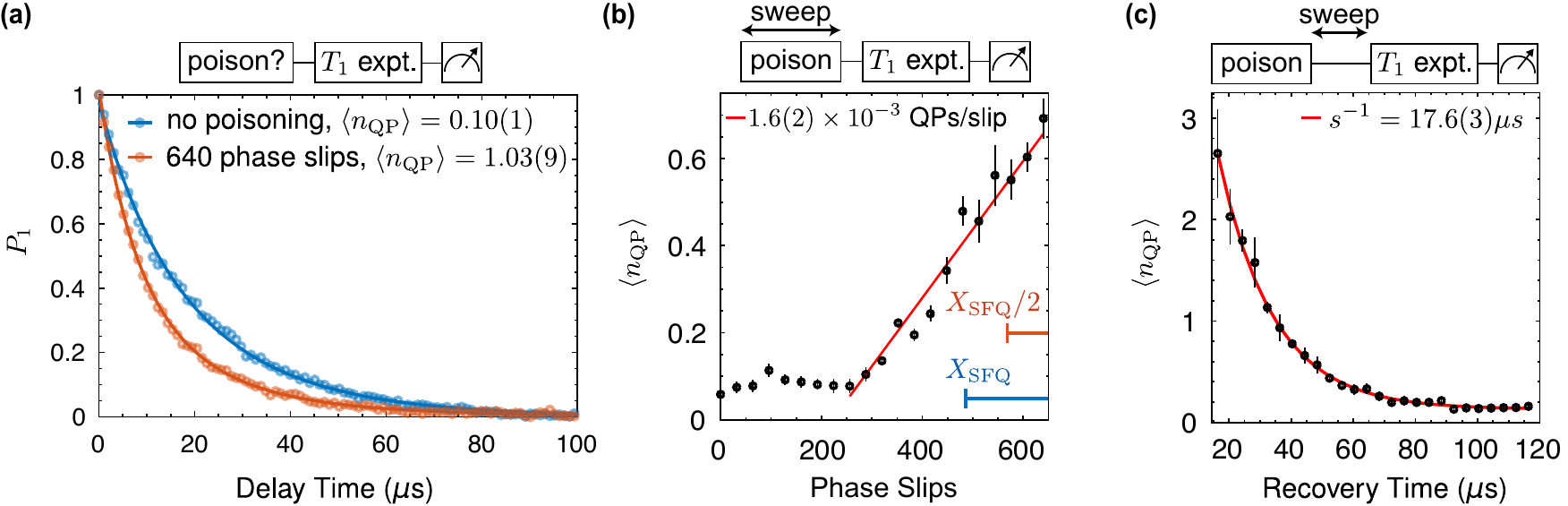}
\caption{\label{fig:qp} (Color) QP poisoning of the qubit induced by operation of the SFQ driver. \textbf{(a)} Energy decay curves with and without a prior applied poisoning pulse from the SFQ driver. The enhanced relaxation rate can be attributed to an increase in the mean number of QPs $\langle n_{{\textrm {QP}}}\rangle$ coupled to the qubit. \textbf{(b)} $\langle n_{{\textrm {QP}}}\rangle$ versus number of phase slips in the SFQ poisoning pulse. We find $1.6(2)\times10^{-3}$ QPs couple to the qubit per JJ phase slip. The lengths of single $X_{\textrm{SFQ}}$ (blue) and $X_{\textrm{SFQ}}/2$ (orange) gates are shown for reference. \textbf{(c)} QP recovery experiment for a fixed poisoning pulse ($\sim20$k phase slips) as a function of the time between the poisoning pulse and the $T_1$ experiment.}
\end{figure*}

With orthogonal control now established, it is possible to generate the full single-qubit Clifford set. We have used interleaved RB~\cite{Knill:2008aa,Magesan:2011aa,Magesan:2012aa} to evaluate the fidelity of SFQ-based Clifford gates. Figure~\ref{fig:rb} shows example depolarizing curves; the table inset lists extracted gate fidelities $\mathcal{F}$ for the $X_{\textrm{SFQ}}, \pm X_{\textrm{SFQ}}/2, Y_{\textrm{SFQ}}, \pm Y_{\textrm{SFQ}}/2$ gates realized with subharmonic drive at both $\omega_{10}/3$ ($X_{\textrm{SFQ}}$-gate time of 28~ns) and $\omega_{10}/41$ ($X_{\textrm{SFQ}}$-gate time of 380~ns). For drive at $\omega_{10}/3$, we find gate fidelities in the range 94-97\%; for drive at $\omega_{10}/41$, the fidelities are only slightly lower, in the range from 91-95\%. In all cases, infidelity is dominated by enhanced qubit relaxation and dispersion induced by QP poisoning from operation of the dissipative SFQ driver; we discuss this effect in detail in the next section. Generally, the shorter $\pi/2$ sequences show higher fidelity than the longer $\pi$ sequences. This can be understood from the fact that the total number of generated QPs depends on the number of phase slips and hence on sequence length. It is also notable that gate fidelities at the two subharmonic drive frequencies are roughly comparable, despite the order-of-magnitude difference in sequence length. This suggests that the improvement in qubit coherence achieved by moving from high to low phase slip (and QP generation) rate is almost exactly compensated by the longer time needed to achieve the desired rotation. The rough balance of the ratio of gate time to coherence time at the two drive frequencies immediately implies that QP removal in the system is dominated by single-particle trapping as opposed to two-particle recombination. In the following, we provide a detailed discussion of QP poisoning in the quantum-classical hybrid circuit.


\section{QUASIPARTICLE POISONING\label{sec:qp}}

Nonequilibrium QPs are a well-known source of decoherence~\cite{Martinis:2009aa,Catelani:2011ab} and temporal instability~\cite{Pop:2014aa,Vool:2014aa,Yan:2016aa} in Josephson qubits. There have been prior detailed studies of QP poisoning in phase~\cite{Lang:2003aa,Lenander:2011aa}, charge~\cite{Lutchyn:2006aa}, flux~\cite{Yan:2016aa}, and transmon~\cite{Riste:2013aa,Wang:2014aa} qubits. A variety of approaches to the suppression of QP poisoning have been explored, including normal metal ``traps''~\cite{Riwar:2016aa,Patel:2017aa,Hosseinkhani:2017aa,Riwar:2018aa}, optimization of device geometry~\cite{Wang:2014aa}, gap engineering~\cite{Aumentado:2004aa,Sun:2012aa}, trapped magnetic flux vortices~\cite{Wang:2014aa,Nsanzineza:2014aa}, and dynamical QP pumping sequences~\cite{Gustavsson:2016aa}. Generally speaking, the most reliable strategy to suppress QP-induced decoherence is to minimize the generation rate of nonequilibrium QPs by housing the qubit in a light-tight enclosure with extensive IR filtering and by employing inline filters in order to minimize the flux of pair-breaking photons to the qubit chip~\cite{Barends:2011aa,Corcoles:2011aa}. For the approach we pursue here, however, it is impossible to avoid QP generation at the device level. Each clock cycle of the trigger waveform induces four phase slips in the driver circuit, one in each JJ of the dc/SFQ converter; each phase slip is accompanied by QP generation. Since the SFQ driver and qubit circuit are located on the same substrate, the generated QPs will poison the qubit circuit either by direct diffusion or by phonon-mediated coupling~\cite{Patel:2017aa}.

We investigate the influence of nonequilibrium QPs by intentionally poisoning the qubit with an off-resonant SFQ drive that generates QPs at a well-defined rate while producing negligible coherent excitation of the qubit. In the experiments described in this Section, all coherent qubit manipulations are performed using conventional microwave techniques, and the SFQ driver is used exclusively to generate QPs. Figure~\ref{fig:qp}\hyperref[fig:qp]{(a)} shows energy relaxation curves for two separate $T_1$ experiments, one with and one without a prior poisoning pulse consisting of 640 phase slips delivered at a rate $\omega_d/2\pi=1.6$~GHz; the poisoning pulse length corresponds to approximately seven $X_{\textrm{SFQ}}/2$ gates. We fit the measured relaxation curves to the following relation to extract the mean number of QPs  coupled to the qubit $\langle n_{\textrm {QP}} \rangle$ by the poisoning pulse~\cite{Pop:2014aa}
\begin{align}
P_1(t) &= \exp\left[{\langle n_{\textrm {QP}}\rangle\left(e^{-t/T_{1,{\textrm {QP}}}}-1\right)-t/T_{1,{\textrm {R}}}}\right];
\label{eqn:nqp}
\end{align}
here, $1/T_{1,{\textrm {QP}}}$ is the decay rate of the qubit per QP and $1/T_{1,{\textrm {R}}}$ is the residual decay rate from all other loss channels. In the absence of an explicit poisoning pulse, we find a background $\langle n_{\textrm {QP}}\rangle = 0.10(1)$ [blue curve in Fig.~\ref{fig:qp}\hyperref[fig:qp]{(a)}]. In contrast, the 640-phase slip poisoning pulse leads to $\langle n_{\textrm {QP}}\rangle=1.03(9)$ [orange curve in Fig.~\ref{fig:qp}\hyperref[fig:qp]{(a)}]. In Fig.~\ref{fig:qp}\hyperref[fig:qp]{(b)} we show results from a series of poisoning experiments where the length of the poisoning pulse is varied prior to the $T_1$ scan. We observe a slow initial turn-on of QP poisoning. For short poisoning pulses, it is likely that nonequilibrium QP density in the vicinity of the SFQ driver is too low to provide for significant phonon-mediated poisoning of the qubit \cite{Patel:2017aa}, as this process relies on high local QP densities to promote recombination and phonon emission in the vicinity of the SFQ driver. However, following approximately 250 phase slips, we observe a linear increase in $\langle n_{\textrm {QP}}\rangle$ with respect to the number of phase slips in the poisoning pulse. In this regime, we find that a single phase slip in the SFQ driver couples approximately $1.6(2)\times10^{-3}$ QPs to the qubit. In a related experiment, we insert a variable delay between a fixed QP poisoning pulse and the $T_1$ scan. In Fig.~\ref{fig:qp}\hyperref[fig:qp]{(c)} we plot $\langle n_{\textrm {QP}}\rangle$ as a function of recovery time following the QP poisoning pulse. We observe an exponential decay of $\langle n_{\textrm {QP}}\rangle$ with an effective trapping time of $s^{-1}=17.6(3)\:\mu$s; this trapping time is consistent with that observed in prior studies of QP poisoning of linear microwave modes in thin-film devices~\cite{Patel:2017aa}.
\begin{figure}[t]
\includegraphics{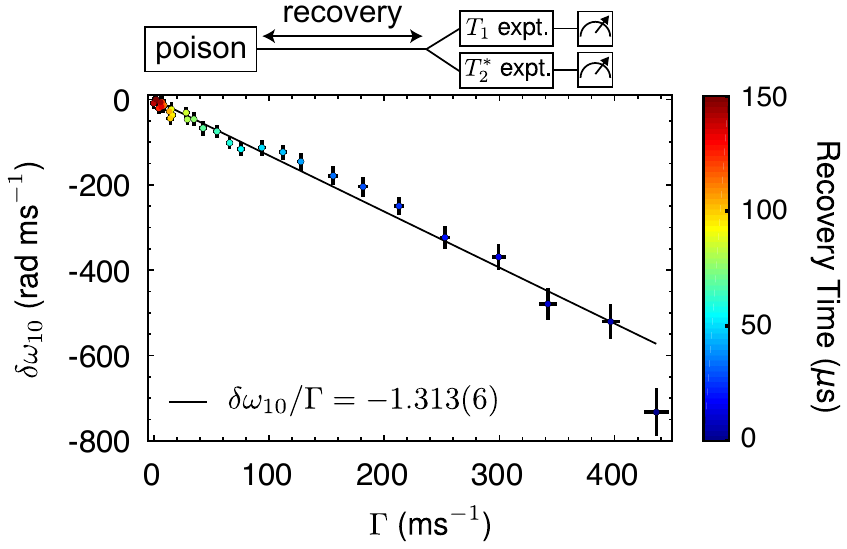}
\caption{\label{fig:shift} (Color) Parametric plot of qubit frequency shift $\delta \omega_{10}$ versus QP-induced decay rate $\Gamma$. Here, QP density at the qubit is controlled by incrementing the recovery time following an intentional QP poisoning pulse prior to the $T_1$ or Ramsey experiment.}
\end{figure}

The complex QP admittance seen by the qubit includes both real and imaginary parts; while the former leads to enhanced relaxation, the latter induces a qubit frequency shift. We have examined the connection between QP loss and dispersion in our device by performing both $T_1$ and Ramsey experiments following intentional QP poisoning pulses with a variable recovery time; these scans provide access to both the QP contribution $\Gamma$ to the qubit relaxation rate and to the frequency shift $\delta \omega_{10}$ induced by the presence of nonequilibrium QPs. In Fig.~\ref{fig:shift} we show a parametric plot of $\delta \omega_{10}$ versus $\Gamma$ for different recovery times; we observe a clear linear relationship between QP loss and dispersion. Within a simplified model that assumes rapid relaxation of nonequilibrium QPs to the gap edge, the ratio of $\delta \omega_{10}$ to $\Gamma$ can be expressed as~\cite{Lenander:2011aa,Catelani:2011ab}
\begin{align}
\frac{\delta\omega_{10}}{\Gamma} &= -\frac{1}{2}\left[1+\pi\sqrt{\frac{\hbar\omega_{10}}{2\Delta}}\right],
\label{eqn:omrate}
\end{align}
where $\omega_{10}$ is the qubit fundamental transition frequency in the absence of QPs and $2\Delta$ is the superconducting gap of the junction electrodes. A fit to our data yields a slope $\delta\omega_{10}/\Gamma$ that is larger than that predicted by Eq.~\hyperref[eqn:omrate]{(}\ref{eqn:omrate}\hyperref[eqn:omrate]{)} by a factor of 1.5; however, the assumption of low-energy QPs is likely invalid in our experiments, and it is expected that this slope will depend sensitively on the details of the QP distribution. 


\section{SUMMARY AND OUTLOOK\label{sec:conc}}

We have described a scheme for coherent qubit control using quantized flux pulses derived from the SFQ digital logic family. The approach offers a path to tight integration of a classical coprocessor with a large-scale multiqubit circuit for the purpose of reducing heat load, latency, and overall system footprint. In our current experiments, we access the full single-qubit Clifford set and achieve gate fidelities $\mathcal{F}\sim95\%$, measured using interleaved RB. This result is limited by QP poisoning induced by operation of the dissipative SFQ pulse driver. However, there are several straightforward modifications that could be employed in a future implementation to significantly suppress QP poisoning. For example, it should be possible to fabricate the classical SFQ driver circuit and the qubit on separate chips that are flip-chip bonded in a multichip module (MCM). With such an MCM arrangement, SFQ pulses would be coupled to the qubit chip capacitively across the chip-to-chip gap of the MCM. The natural choice for the MCM bump bonds, indium, would provide a low-gap barrier to direct QP diffusion between the chips, and recombination phonons generated by QPs in the bumps would have insufficient energy to break pairs in the niobium groundplane of the quantum chip. The MCM approach would have the additional advantage of allowing a modular fabrication, obviating the need to develop a process flow that protects the delicate quantum circuit from materials-induced loss associated with the complex, multilayer SFQ driver stack. Ultimately, we view the MCM approach as the most promising path to scaling the two-dimensional surface code~\cite{Fowler:2012aa}, as it provides for the necessary nearest-neighbor connectivity of the quantum array while allowing tight integration of proximal control and measurement elements. There has been significant recent progress in the development of robust superconducting indium bump bonding of MCMs~\cite{Rosenberg:2017aa,OBrien:2017aa,Liu:2017aa,Bronn:2018aa,Foxen:2018aa}, and it has been shown that qubit coherence can be fully preserved in a properly designed MCM~\cite{Rosenberg:2017aa}. We have previously outlined an approach to scaling superconducting qubits based on MCMs incorporating an interface chip that houses both SFQ control and measurement elements~\cite{McDermott:2018aa}.

A second straightforward modification to suppress QP poisoning would be to move from a high-$J_c$ process for the SFQ driver circuit to a low-$J_c$, millikelvin-optimized process. SFQ control elements should remain robust against fluctuations at millikelvin temperatures for critical currents that are two orders of magnitude smaller than those used in the current work~\cite{Castellano:2006aa}; as the energy dissipated per phase slip scales linearly with critical current, a millikelvin-optimized SFQ driver is expected to generate nonequilibrium QPs at a factor $\sim$100 lower rate. The inductances required for such an SFQ pulse driver would be correspondingly larger, leading to a larger physical footprint for the SFQ controller; global optimization of the SFQ control system would require tradeoffs between dissipation and physical size. 

Finally, the incorporation of normal metal QP traps in the hybrid SFQ-qubit circuit or MCM is expected to significantly suppress QP poisoning. When QPs diffuse into a normal metal, they interact strongly with electrons and quickly relax below the gap edge, so that they are unable to re-enter the superconductor. The effectiveness of normal metal traps in removing nonequilibrium QPs has been demonstrated previously in a variety of contexts~\cite{Ullom:2000aa,Pekola:2000aa,Giazotto:2006aa,Muhonen:2012aa,Riwar:2016aa}. In related work, it has been shown that appropriate engineering of the spatial profile of the superconducting gap in single-Cooper-pair transistor (SCPT) devices can be used to confine nonequilibrium QPs to regions of the circuit where they are unable to degrade device performance~\cite{Aumentado:2004aa}. For the application we consider here, the generation of QPs is localized to the driver elements that undergo phase slips; the phase slips lead to an elevated local population of nonequilibrium QPs in the vicinity of the driver. Poisoning proceeds by multiple stages of phonon emission and pair-breaking, so that extensive coverage of the device with normal metal traps (as opposed to local coverage near or surrounding the QP generation point) is required to suppress poisoning. The addition of QP traps has been shown to suppress QP poisoning of linear resonator modes by 1-2 orders of magnitude~\cite{Patel:2017aa}. The optimal configuration of QP traps in a quantum-classical MCM is a question that would likely need to be addressed experimentally.

Once the problem of QP poisoning is addressed, we expect leakage out of the computational subspace to represent the dominant source of error in SFQ-based gates. While naive, resonant pulse sequences of the type explored here are expected to yield gate fidelity of order 99.9\%~\cite{McDermott:2014aa}, it has been shown that more complex SFQ pulse sequences involving variable pulse-to-pulse intervals can yield gates with fidelity better than 99.99\%~\cite{Liebermann:2016aa}, compatible with the fidelities of the best reported microwave-based gates and sufficient for scaling in the two-dimensional surface code.


\section*{ACKNOWLEDGEMENTS}

The authors thank O. A. Mukhanov, T. A. Ohki, M. J. Vinje, and F. K. Wilhelm for numerous stimulating discussions. This work was supported by the U.S. Government under Grant W911NF-15-1-0248; R.M. and B.L.T.P. acknowledge funding from the National Science Foundation under Grants QIS-1720304 and QIS-1720312, respectively. Portions of this work were performed at the Wisconsin Center for Applied Microelectronics, a research core facility managed by the College of Engineering Shared Research Facilities and supported by the University of Wisconsin - Madison. Other portions were performed at the Cornell NanoScale Facility, a member of the National Nanotechnology Coordinated Infrastructure (NNCI), which is supported by the National Science Foundation under Grant No. ECCS-1542081. Microscopy and materials analysis infrastructure and instrumentation used for this work were supported by the National Science Foundation through the University of Wisconsin Materials Research Science and Engineering Center (DMR-1720415). The views and conclusions contained in this document are those of the authors and should not be interpreted as representing the official policies, either expressed or implied, of the U.S. Government.


\appendix
\titleformat{\section}{\centering\bfseries\small}{\MakeUppercase{\appendixname~\thesection .}}{0.5em}{}

\section{FABRICATION DETAILS\label{app:fab}}

\begin{table*}[!ht]
\begin{tabular}{ccccccccl}
\thead{Layer\\ID} & \thead{Material} & \thead{Thickness \\ (nm)} & \thead{Surface \\ Pre-treatment} & \thead{Deposition \\ Method} & \thead{Patterning} & \thead{Etch \\ Chemistry} & \hspace{2 pt} &\thead[l]{Function}  \\
\hline\hline
\makecell{M1\\~\\~ } &\makecell{Nb\\~\\~ } & \makecell{180\\~\\~ } & \makecell{hydrofluoric acid\\~\\~ } & \makecell{dc sputtered\\~\\~ } & \makecell{i-line\\~\\~ } & \makecell{\chem{Cl_2/BCl_3}\\~\\~ } &  &\makecell[l]{ground plane \\ resonators \\ capacitors}\\
\hline
\makecell{D1\\~} & \makecell{\chem{SiO_x}\\~} & \makecell{130\\~} & \makecell{--\\~} & \makecell{PECVD\\~} & \makecell{i-line\\~} & \makecell{\chem{CHF_3}\\~} & &\makecell[l]{SFQ circuit wiring \\ ground vias} \\
\hline
\makecell{M2\\~} & \makecell{Nb\\~} & \makecell{90\\~} & \makecell{\textit{in situ} ion mill\\~}& \makecell{dc sputtered\\~} & \makecell{i-line\\~} & \makecell{\chem{SF_6}\\~} & &\makecell[l]{inductors \\ SFQ JJ base layer}\\
\hline
\makecell{D2\\~} & \makecell{\chem{SiO_x}\\~} & \makecell{180\\~} & \makecell{--\\~} & \makecell{PECVD\\~} & \makecell{i-line\\~} & \makecell{\chem{CHF_3}\\~} & &\makecell[l]{junction definition \\ shunt resistor vias} \\
\hline
\makecell{M3\\~\\~} & \makecell{Nb\\Al-\chem{AlO_x}-Al\\~} & \makecell{100\\~\\~} & \makecell{\textit{in situ} ion mill\\~\\~}& \makecell{dc sputtered \\ w/\textit{in situ} oxidation\\~} & \makecell{i-line\\~\\~} & \makecell{\chem{SF_6}\\\chem{TMAH}\\~} & &\makecell[l]{SFQ bias wiring \\ SFQ JJ counterelectrodes \\ CPW crossovers}\\
\hline
R & Ti/Pd & 23 & \textit{in situ} ion mill& e-beam evaporated & i-line & -- & &shunt resistors\\
\hline
\makecell{QB\\~} & \makecell{Al-\chem{AlO_x}-Al\\~}  & \makecell{100\\~} & \makecell{\textit{in situ} ion mill\\~}& \makecell{e-beam evaporated \\ (double-angle)} & \makecell{e-writing \\ (Dolan Bridge)} & -- & &\makecell[l]{qubit JJs\\~}\\
\hline
\end{tabular}
\caption{\label{tab:fab} Layer stack of the hybrid quantum-classical circuit.}
\end{table*}

The dc/SFQ converter and transmon qubit are co-fabricated in a multilayer process on a high resistivity ($>20\mbox{ kOhm-cm}$)~Si$(100)$ wafer. Fabrication is realized in the following steps, which are also summarized in Table~\ref{tab:fab}:

\begin{figure}[b]
\includegraphics[width=0.48\textwidth]{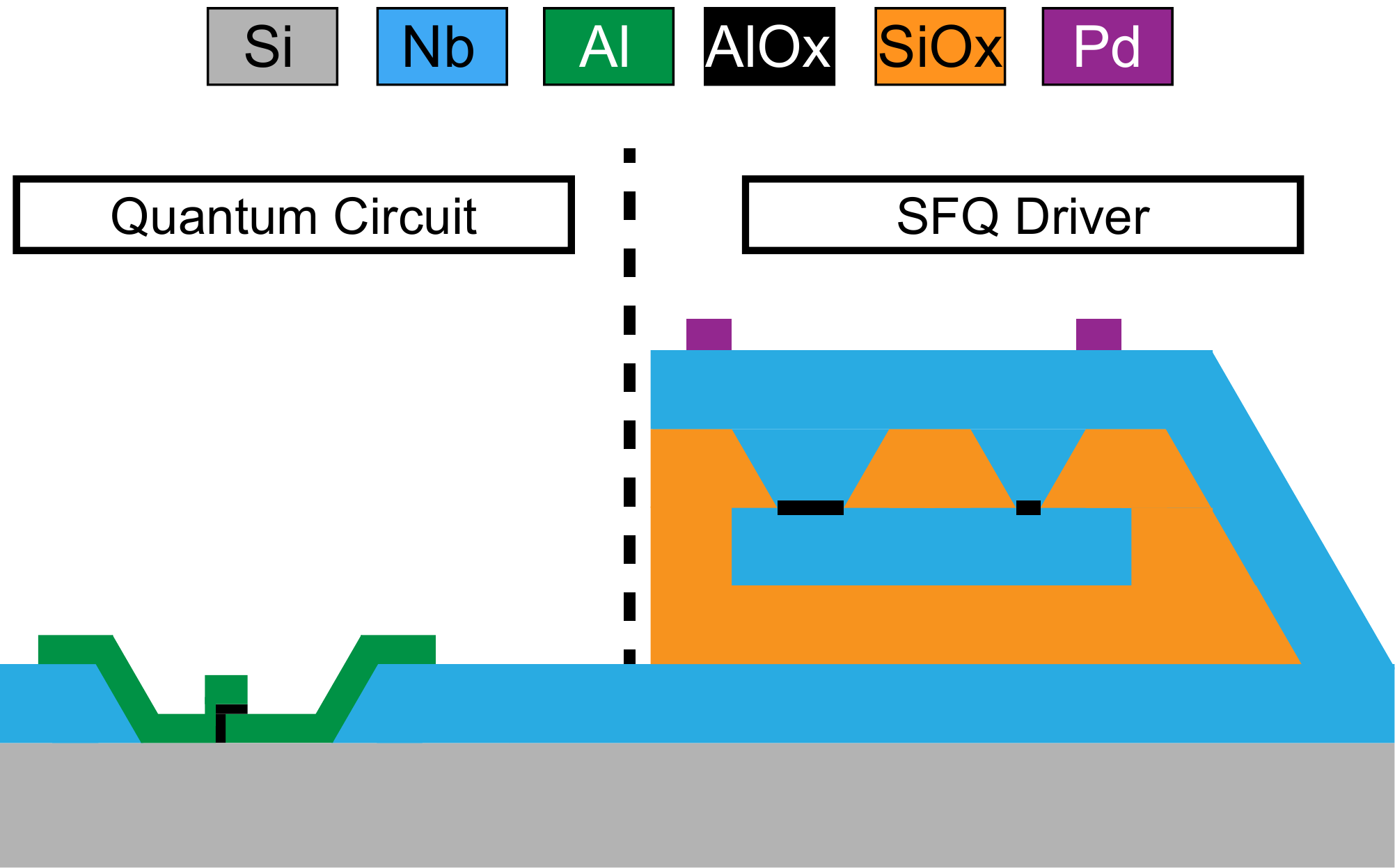}
\caption{\label{fig:resonator} (Color) Device layer stack described in Appendix~\ref{app:fab}. During the entirety of the fabrication process of the SFQ driver circuit, the area that will support the quantum circuit is left unpatterned and unetched and is protected by the first deposited layer of \chem{SiO_x}.}
\end{figure}

\begin{figure*}[h]
\includegraphics[width=17.8cm]{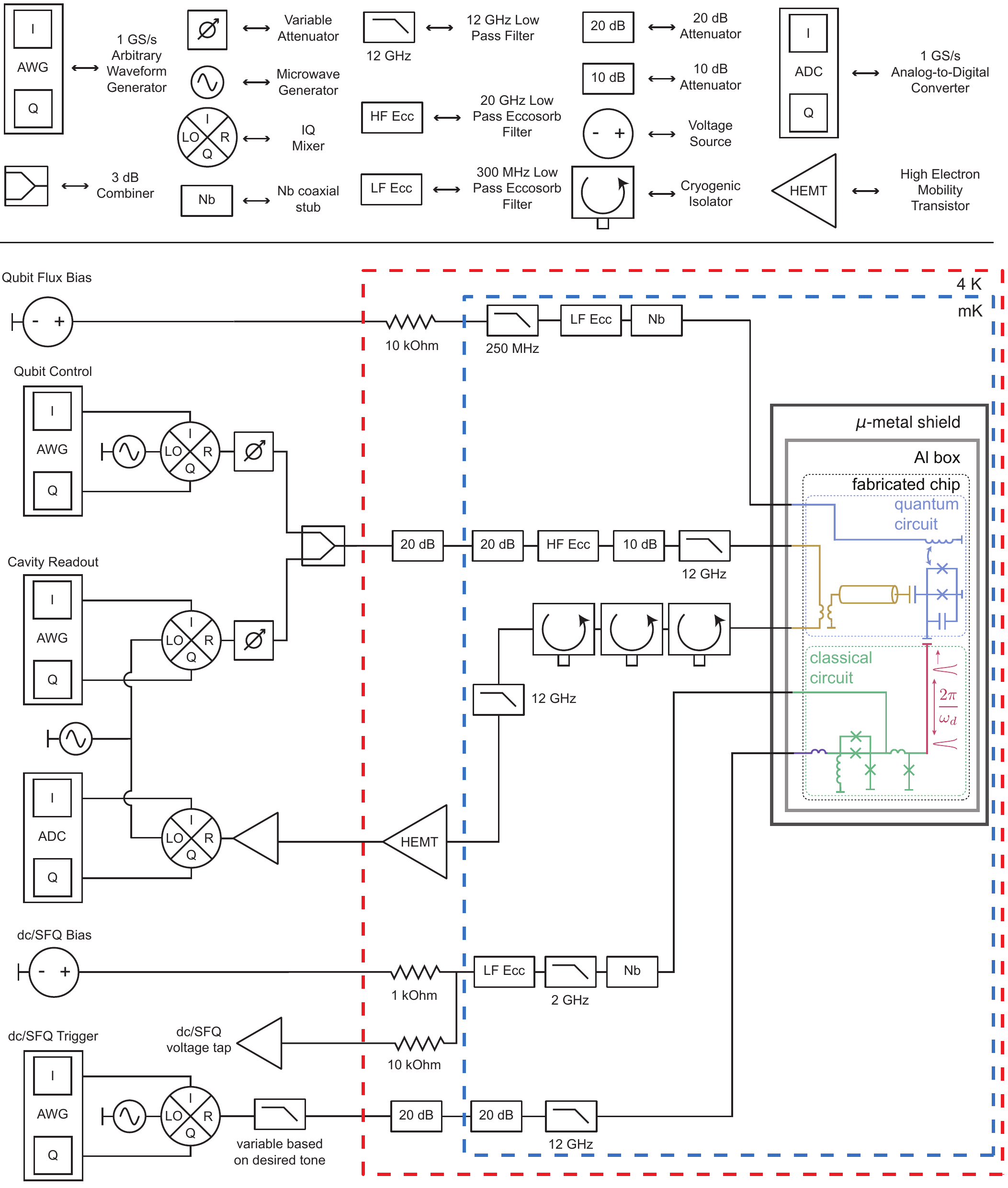}
\caption{\label{fig:wiring} Experimental wiring diagram. Component symbols are defined in the panel above the wiring diagram. All components within the $\mu$-metal shield (grey box) were made of either non-ferric or certified non-magnetic materials.}
\end{figure*}

\begin{enumerate}
\item \label{step:sputter} A bare intrinsic Si wafer is coated with a film of Nb (M1) after stripping the native \chem{SiO_x} with hydrofluoric (HF) acid. Here and below, sputter conditions are tuned to yield films with slight compressive stress~\cite{Kuroda:1988aa}; the deposition rate is 45~nm/min. This layer is patterned with an i-line step-and-repeat (stepper) lithography tool and etched with a \chem{Cl_2}/\chem{BCl_3}-based inductively coupled plasma (ICP). During this and the following six steps, the area of the die that will ultimately support the readout resonator and qubit is left unpatterned and unetched.

\item \label{step:v1} A plasma-enhanced chemical vapor deposition (PECVD) process is used to grow a conformal layer of \chem{SiO_x} (D1). This layer is patterned with an i-line stepper and etched with a \chem{CHF_3}-based reactive ion etch (RIE) plasma. The RIE power and pressure are optimized to transfer a $45^{\circ}$ slope into the dielectric sidewalls in order to promote step coverage of subsequent layers [see Fig.~\ref{fig:overview}\hyperref[fig:overview]{(c)}]. 

\item The bottom Nb electrode (M2) for the SFQ JJs is sputtered. This layer is patterned with an i-line stepper and etched with an \chem{SF_6}-based RIE plasma.

\item Another \chem{SiO_x} layer (D2) is grown by PECVD. This layer is patterned in the same manner as in step~\ref{step:v1}. At this step, particular care is taken to eliminate solvent or other organic residues from the dielectric vias, as these vias will define the SFQ JJs.

\item{The SFQ JJs and bias wiring are formed in the following process (M3):
  \begin{enumerate}[label=(\alph*)]
  \item The wafer is transferred into the sputter system load-lock chamber and the chamber is pumped down for $\sim1$~hour.
  \item While the load lock is pumping out, the primary vacuum chamber is ``seeded'' with pure \chem{O_2} using similar conditions to the actual junction oxidation. We find that this step improves the reproducibility of the junction specific resistance by ensuring as similar a chamber chemistry as possible from run to run.
  \item The wafer is transferred into the primary vacuum chamber and the system is allowed to pump down to base pressure (about $5\times10^{-8}$~Torr).
  \item An \textit{in situ} ion mill is used to remove the \chem{NbO_x} from the M2 electrode layer.
  \item An Al seed layer of thickness $\sim8$~nm is sputtered.
  \item The barrier oxide is grown. The \chem{AlO_x} barrier layer is seeded by flowing \chem{O_2} at low pressure ($\sim$ 1~mTorr) for 2~minutes. The chamber is then valved off from the pumps and the \chem{O_2} pressure is quickly ramped to the target value (typically $\sim100$~mTorr), where it remains for the duration of the oxide growth (typically about 10 minutes).
  \item The growth chamber is pumped back to vacuum. An Al cap layer of $\sim6$~nm thickness is sputtered, followed by the 90~nm Nb junction counterelectrode.
  \end{enumerate}
  This layer is patterned with an i-line stepper. The Nb is removed with an \chem{SF_6}-based RIE plasma while the Al-\chem{AlO_x}-Al is removed with a TMAH-based photoresist developer wet etch.}

\item \label{step:probe}At this point, it is possible to probe the 4-wire resistances of witness junctions cofabricated on the die with the SFQ/qubit circuit. The expected critical currents are extracted using the Ambegaokar-Baratoff relation~\cite{Ambegaokar:1963aa}.

\item The \chem{SiO_x} layer protecting the area of the qubit and readout resonator is now patterned and etched as in step~\ref{step:v1}.

\item The readout resonators and qubit capacitor are defined in the M1 Nb layer using pattern and etch processes as in step~\ref{step:sputter}.

\item A negative photoresist is patterned with an i-line stepper to define the shunt resistors (R). The resistors are formed by electron-beam evaporation of Pd following an \textit{in situ} ion mill and deposition of a thin ($\sim$3~nm) Ti adhesion layer. Liftoff is performed in acetone at room temperature over several hours with slight manual agitation of the wafer.

\item \label{step:ejj}The qubit junctions are defined using a Dolan bridge process~\cite{Dolan:1977aa} involving an MMA/PMMA stack patterned with a 100 keV electron-beam writer. The Al-\chem{AlO_x}-Al stack is shadow evaporated in a high vacuum electron beam evaporation tool following an \textit{in situ} ion mill to ensure good metallic contact to the base layer (M1).
\end{enumerate}

\section{MEASUREMENT SETUP\label{app:meas}}

The measurement setup is shown in Fig.~\ref{fig:wiring}. The microwave tones used to read out the qubit, perform microwave control of the qubit, and trigger the SFQ pulse driver are all generated through single-sideband modulation of a local oscillator (LO) carrier wave with shaped intermediate frequency (IF) signals. Arbitrary waveform generators (AWG) with an output rate of 1~GS/s and 14-bit resolution generate the IF tones and are directly connected to the in-phase (I) and quadrature (Q) ports of an IQ mixer. The shaped microwave pulses for qubit control and readout are passed through digital variable attenuators before merging on a 3~dB combiner and entering the cryostat. The readout and microwave control tones pass through multiple stages of attenuation and filtering at the 4~K and mK stages of the dilution refrigerator prior to reaching the device. After interacting with the device, the qubit readout tone passes through several stages of isolation and filtering prior to amplification at 4~K using a high electron mobility transistor (HEMT) amplifier. The readout tone is further amplified at room temperature. The same LO used to generate the readout signal is used for heterodyne detection of the qubit state. The downconverted IF tones from qubit measurement are sampled at 500~MS/s with 8-bit resolution and demodulated and analyzed in software. For all of the work presented here, we employ so-called ``bright-state'' readout of the qubit~\cite{Reed:2010aa}. The microwave tone used to trigger the SFQ pulse driver is conditioned with a low-pass filter that provides strong rejection at the qubit fundamental frequency $\omega_{10}$. The trigger tone passes through multiple stages of attenuation and filtering at the 4~K and mK stages of the cryostat before reaching the device. Both the qubit flux bias and the dc bias current for the SFQ pulse driver are generated by isolated low-noise voltage sources at room temperature. These signals are conditioned by multiple stages of filtering at the 4~K and mK stages of the cryostat.

\bibliographystyle{apsrev4-1-new}
\bibliography{master_refs}

\end{document}